\documentclass[letterpaper, 10 pt, conference]{ieeeconf}

\IEEEoverridecommandlockouts                              %

\usepackage{amssymb}

\usepackage{subcaption}

\usepackage{comment}
\usepackage{booktabs}

\usepackage{graphicx}
\usepackage{amsmath}
\usepackage{verbatim}
\usepackage{amsfonts}
\usepackage{multirow}
\usepackage{changepage}
\usepackage{tabularx}
\usepackage{adjustbox}
\usepackage{mathtools}
\usepackage{amssymb}
\usepackage{soul}
\usepackage{color}
\usepackage[noend]{algpseudocode}

\usepackage{color}
\usepackage{tikz}
\usepackage{flushend}
\usetikzlibrary{shapes.geometric, arrows}
\usepackage[linesnumbered,ruled]{algorithm2e}
\usepackage{multirow}
\usepackage{soul,color}
\usepackage{graphicx}
\usepackage{fancyhdr}
\usepackage{comment}
\usepackage{epsfig}
\usepackage{bm}
\usepackage{float}
\usepackage{cuted}
\usepackage{url}

\usepackage{makecell}

\newcommand\oprocendsymbol{\hbox{$\square$}}
\newcommand\oprocend{\relax\ifmmode\else\unskip\hfill\fi\oprocendsymbol}

\newtheorem{remark}{Remark}

\renewcommand{\theenumi}{(\roman{enumi})}

\allowdisplaybreaks

\begin{document}


\title{Optimal Fidelity Selection for Human-Supervised Search    
\thanks{This work has been supported in part by the NSF awards IIS-$1734272$, ECCS-$2024649$, and the ONR award N$00014$-$22$-$1$-$2813$.}
}

\author{Piyush Gupta \hspace{2cm}  Vaibhav Srivastava 
 \thanks{P. Gupta (guptapi1@msu.edu) and V. Srivastava (vaibhav@egr.msu.edu) are with Department of Electrical and Computer Engineering, Michigan State University, East Lansing, Michigan, 48824, USA.}
 }

\maketitle

\begin{abstract}
We study optimal fidelity selection in human-supervised underwater visual search, where operator performance is affected by cognitive factors like workload and fatigue. In our experiments, participants perform two simultaneous tasks: detecting underwater mines in videos (primary) and responding to a visual cue to estimate workload (secondary). Videos arrive as a Poisson process and queue for review, with the operator choosing between normal fidelity (faster playback) and high fidelity. Rewards are based on detection accuracy, while penalties depend on queue length. Workload is modeled as a hidden state using an Input-Output Hidden Markov Model, and fidelity selection is optimized via a Partially Observable Markov Decision Process. We evaluate two setups: fidelity-only selection and a version allowing task delegation to automation to maintain queue stability. Our approach improves performance by 26.5\% without delegation and 50.3\% with delegation, compared to a baseline where humans manually choose their fidelity levels. 

\end{abstract}

\section{Introduction}

Human-in-the-loop systems are extensively used in domains like search and rescue, security screening, and tele-operated robotics~\cite{amador2022survey, gupta2022incentivizing, wang2023human}. These systems often place high cognitive demands on humans, leading to fatigue and errors due to visual limitations and attention bottlenecks. For example, in underwater search tasks, reviewing long sequences of images can increase workload, raising the risk of missed detections or false alarms. Consequently, managing operator workload is critical for maintaining performance and minimizing errors.


We study optimal fidelity selection~\cite{gupta2019optimal, 9654941} for a human operator performing a visual search task, with fidelity referring to the level of precision required for task completion. Focusing on underwater mine detection from simulated videos, we model video arrivals as a Poisson process queued for operator review. The operator chooses between normal fidelity (faster but less precise) and high fidelity (slower but more accurate), creating a trade-off between processing speed and detection accuracy. We analyze this trade-off to derive an optimal fidelity selection policy.


In addition to fidelity, task performance is also influenced by operator workload~\cite{ jain2023enabling, byeon2025workload}, with high workload increasing the error likelihood, especially under normal fidelity. Thus, an effective fidelity selection policy must account for both queue length and operator workload to optimize performance. Similar class of problems, such as fidelity selection and attention allocation under workload dynamics, have been explored in prior work~\cite{KS-EF:10b, PG-VS:21m, VS-RC-CL-FB:11zc} and apply broadly, e.g. in domains like airport security screening. However, experimental validation is limited~\cite{cummings2006automated, crandall2010computing}, partly due to the difficulty in measuring the workload state.



Human workload is often estimated indirectly using physiological sensors 
that capture features like heart rate, pupil diameter, and blink rates to infer cognitive states like fatigue or situational awareness~\cite{heard2018survey}. However, these signals can reflect more than just workload and often require costly hardware, limiting their practicality, especially for remote supervision. As a more accessible alternative, workload can be estimated using the operator’s reaction time to a simple secondary task, consistent with prior research~\cite{akash2019improving, helldin2014transparency}.

Recent studies have focused on designing efficient human-in-the-loop systems that integrate human knowledge with autonomy to manage cognitive resources effectively~\cite{gupta2023optimal, feng2016synthesis}. Optimal fidelity selection was studied in~\cite{PG-VS:21m}, which established the structural properties of the optimal policy under known cognitive dynamics modeled as a Markov chain. Other works include optimal task scheduling~\cite{peters2018robust, kaza2025task}, operator allocation for multi-robot
assistance~\cite{dahiya2022scalable}, and optimal attention allocation~\cite{yao2024optimal}. In contrast, we assume the human workload is a hidden state and experimentally evaluate the impact of optimal fidelity selection on performance.

We model the human workload as a hidden state using an Input-Output Hidden Markov Model (IOHMM)~\cite{10156103}, where input fidelity affects workload dynamics. This hidden state impacts the operator's performance in primary and secondary tasks, which we use as observations to estimate the workload.  Alternatively, similar to secondary tasks, other noisy workload measures, such as blink rate, can also be incorporated into our framework. Based on the learned state and observation dynamics, a Partially Observable Markov Decision Process (POMDP)~\cite{braziunas2003pomdp} is employed to derive an optimal fidelity selection policy. Experiments reveal that this optimal policy significantly improves human performance compared to self-selected policies.


This work has three major contributions. First, we address the optimal fidelity selection challenge by framing it as a control of a queue problem, with a hidden server workload state. We employ IOHMM and POMDP to derive the optimal fidelity selection policy. Second, we compare the human fidelity selection policy with the optimal policy and draw valuable insights into human behavioral patterns. Third, we illustrate that by recommending the optimal policy, a decision support system can effectively aid human decision-making, leading to a substantial improvement in their performance.

The rest of the manuscript is organized as follows. Section~\ref{ch_hil:sec:background} presents our problem setup and mathematical model. Section~\ref{ch_hil:sec:human_experiments} details the design of our human experiments. Section~\ref{sec:resultsanddiscussion} discusses our primary findings, including a comparison between human and optimal policies. Finally, Section~\ref{sec:conclusionsandlimitations} concludes with a discussion of our work's limitations.

\section{Background and Problem Formulation}\label{ch_hil:sec:background}
\begin{figure}
\centering
 \includegraphics[width=0.7\linewidth, height=\linewidth, keepaspectratio]{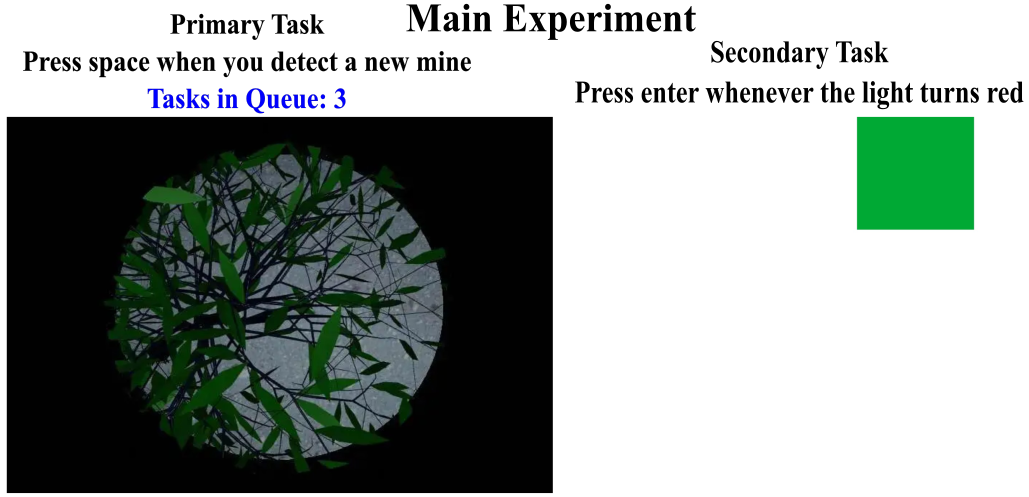}
 \caption{\footnotesize{Human experiment interface. The participants press the spacebar whenever a new mine is detected in the primary task video. Additionally, the green light (secondary task) randomly turns red once for each primary task and the participant responds by pressing the Enter key as early as possible. The queue length is displayed on top of the primary task.}}
    \label{ch_hil:experiment_task}
\end{figure}
In this section, we formulate the problem as a POMDP and solve for the optimal fidelity selection policy.
\subsection{Problem Setup}
We study optimal fidelity selection for a human-supervised visual search task involving primary and secondary tasks. The primary task involves searching for underwater mines in videos from a Gazebo and ROS-based simulation, with the operator pressing a key upon spotting a mine. The secondary task requires participants to press a key when a green light turns red, with the color change occurring randomly between $25\%$ and $75\%$ of the video. These thresholds are chosen to intentionally avoid the beginning and end of the video, reducing predictability and boundary effects.
Reaction times for the secondary task are recorded, and if a participant misses the red light, the reaction time is set to the total duration the light stays red until the end of the primary task.

Fig.~\ref{ch_hil:experiment_task} shows the experiment interface. The primary task videos arrive as a Poisson process with an arrival rate $\lambda \in \mathbb{R}_{>0}$ and get stacked in a queue awaiting service by the human operator. The operator selects either high or normal fidelity levels for servicing each video. In normal fidelity, the video is presented at a speed of three times faster compared to high-fidelity processing. Additionally, the operator can choose to delegate a task for autonomous processing, which may have lower accuracy.
This delegation or ``skip" action serves as a means to maintain queue stability, especially in situations with large queue lengths.

\begin{figure}
\centering
 \includegraphics[width=0.5\linewidth, height=\linewidth, keepaspectratio]{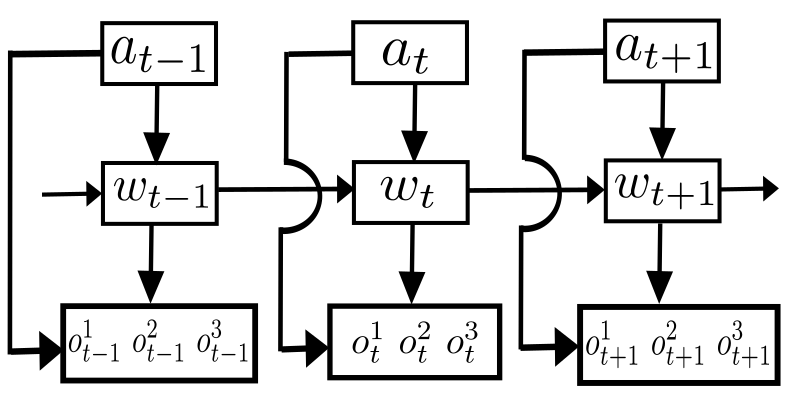}
 \caption{\footnotesize{Input-output hidden Markov model. The input $a$ represents the fidelity level, the hidden state $w$ signifies the workload, and the three observations $o^1$, $o^2$, and $o^3$ correspond to the fraction of correctly detected mines in the primary task, the count of false alarms in the primary task, and the reaction time recorded in the secondary task, respectively.}}
    \label{ch_hil:IOHMM_layout}
\end{figure}

The human operator’s performance is influenced by their workload, which we treat as a hidden variable.
We estimate this workload using reaction times from the secondary task and performance metrics from the primary task. Specifically, we use an IOHMM (see Fig.~\ref{ch_hil:IOHMM_layout}) where the fidelity level $a$ is the input, workload $w$ is the hidden state, and the outputs 
$o^1, o^2, o^3$ represent, respectively, the fraction of correctly detected mines, the number of false alarms in the primary task, and the reaction time during the secondary task. This framework links fidelity, workload, and task performance in a unified model.

We train the IOHMM using the extended Baum-Welch algorithm~\cite{bengio1996input} to compute the transition probabilities $p(w' | w, a)$, observation probabilities $p(o | w, a)$, and the initial state distribution $p(w_0)$ via expectation maximization. These probabilities are then used in the POMDP formulation to derive the optimal fidelity selection policy.
 

We determine the optimal number of hidden workload states using the Akaike information criterion (AIC) and Bayesian information criterion (BIC)~\cite{burnham2004multimodel}. For a given number of hidden states, the AIC and BIC are defined as:
\begin{equation}
    AIC = 2p - 2\log(\hat{\mathcal{L}}), \ \ \ BIC = p\log(n_o) - 2\log(\hat{\mathcal{L}}),
\end{equation}
where $p$ is the number of learned parameters, $n_o$ is the number of observation trajectories, and $\hat{\mathcal{L}}$ is the maximized log-likelihood of the trained model. The model with the lowest AIC or BIC value is selected as optimal. Based on these criteria (detailed in Section~\ref{subsec:IOHMM}), we train an IOHMM model with two hidden states,  which we refer to as normal and high workload states.

\begin{remark}
    We assume that the human workload remains stationary during the processing of a single video task. This is motivated by the short duration of each video ($5–15$ seconds), which is typically insufficient to induce substantial intra-task fluctuations in cognitive workload. Moreover, modeling within-task workload transitions would require richer observational granularity (e.g., sub-task segmentation), which is not feasible in the current experimental design.
\end{remark}

\subsection{Mathematical Modeling}\label{ch_hil:subsec:mathematical_modeling}
We formulate our problem as a POMDP $\mathcal{P} = \{\mathcal{S},$ $\mathcal{A},$ $\Omega,$ $\mathcal{T},$ $\mathcal{O},$ $r,$ $\gamma \}$, where $\mathcal{S}$ is the state space, $\mathcal{A}$ in the action space,  $\Omega$ is the set of observations, $\mathcal{T}$ is the set of conditional transition probabilities between states, $\mathcal{O}$ is the set of conditional observation probabilities, $r: \mathcal{S}\times \mathcal{A} \rightarrow \mathbb{R}$ is the reward function, and $\gamma \in [0, 1)$ is the discount factor. The state space is defined as $\mathcal{S} = (q, w)$, where $q \in \{0, 1, \ldots, L\}$ and $w \in \mathcal{W} = \{0, 1\}$ denote the number of tasks waiting in the queue (with maximum queue length $L$) and the hidden discrete human workload, respectively. We define $w=0$ as the normal workload state and $w=1$ as the high workload state. The action space $\mathcal{A} = \{N, H, D\}$ consists of normal ($N$) and high ($H$) fidelity processing, along with delegation ($D$), which offloads tasks to an autonomous system and is useful for maintaining queue stability when dealing with large queue lengths. To discourage overuse of $D$, autonomy’s mine-detection accuracy is assumed lower, yielding a reduced immediate reward.

 The observation space is defined as $\Omega = (o^1, o^2, o^3)$, where $o^1$, $o^2$, and $o^3$ correspond to the fraction of correctly detected mines, the count of false alarms in mine detection, and the reaction time recorded during the secondary task, respectively. The state transition probability $p(s'|s,a) \in \mathcal{T}$ is derived from the queue dynamics $p(q'|q, a)$ given by the Poisson distribution with arrival rate $\lambda t$, where $t$ is the duration of the task, and workload dynamics $p(w'|w, a)$ obtained by training the IOHMM. The observation probabilities  $p(o|s',a) = p(o^1,o^2,o^3|w',a) \in \mathcal{O}$ are also obtained from the trained IOHMM model. The observation probabilities are assumed to be independent of the queue length, and therefore, only depend on the operator workload, i.e., $p(o|s',a) = p(o|w',a)$. 
 We define the reward function as $r(s,a) = \alpha_1 o^1 - \alpha_2 o^2 -\alpha_3 q$, where $\alpha_i$ for $i \in \{1,2,3\}$ are positive constants, which rewards high accuracy in the primary task and penalizes for the number of tasks waiting in the queue. The values of $\alpha_i$ chosen for the empirical study are provided in \eqref{ch:hil:eq:reward}.

 We convert the POMDP to a belief MDP defined by $\mathcal{M} = \{\mathcal{B}, \mathcal{A}, \tau, r, \gamma\}$. Here, $\mathcal{B}:= \{(q,b_H) | \  q \in \{0, 1, \ldots, L\}, b_H \in \Delta_D \}$ is the new state space, where $q$ is the original queue length, $b_H$ is the discrete belief probability for being in the high workload state, and $\Delta_D$ is a discretization of the interval $[0,1]$. Therefore, the belief probability for being in the normal workload state is given by $1-b_H$. For our experiments, we discretize $[0,1]$ with a step size of $0.1$, i.e., $\Delta_D = \{0, 0.1, \ldots, 1\}$. Note that the discretization of $b_H$ results in a finite state space $\mathcal{B}$. Let $b : \mathcal{W} \rightarrow \Delta_D$ denote the belief vector, where $b(0) = 1-b_H$, and $b(1) = b_H$.     
    From a current belief $b(w)$, taking an action $a$ and observing $o$, the updated belief $b'(w')$ is given by:
\begin{equation}\label{ch_hil:eq:belief_update}
    b'(w') = \eta p(o|w',a)\sum_{w \in \mathcal{W}}p(w'|w,a)b(w),
\end{equation}
where $\eta = \frac{1}{p(o|b,a)}$ is the normalizing constant with 
\begin{equation}\label{ch_hil:eq:p_o_b_a}
    p(o|b,a) = \sum_{w' \in \mathcal{W}}p(o|w',a))\sum_{w \in \mathcal{W}}p(w'|w,a)b(w).
\end{equation}
The updated belief probabilities in $b'(w')$, where $b'(0) = 1- b'_H$ and $b'(1) = b'_H$ obtained from~\eqref{ch_hil:eq:belief_update} are mapped to the closest discrete states such that $b'_H \in \Delta_D$. The action set $\mathcal{A}$ is the original action space. The transition probabilities $\tau(q',b'|q,b,a) = p(q'|q,a)p(b'|b,a)$ is composed of the Poisson process for queue dynamics $p(q'|q,a)$, and the workload dynamics $p(b'|b,a)$, which is given by:
\begin{equation}
    p(b'|b,a) = \sum_{o \in \mathcal{O}}p(b'|b,a,o)p(o|b, a), \ \text{where}
\end{equation}
\begin{equation}
    p(b' | b, a, o) = \begin{cases}
        1, & \text{if belief update in~\eqref{ch_hil:eq:belief_update} returns $b'$}, \\
        0, & \text{otherwise,}
    \end{cases}
\end{equation}
and $p(o|b, a)$ is defined in~\eqref{ch_hil:eq:p_o_b_a}. The reward function $r : \mathcal{B}\times \mathcal{A} \rightarrow \mathbb{R}$ is given by:
\begin{equation}
            r(q, b_H,a) = \sum_{s \in \mathcal{S}| w = 0}r(s,a)(1-b_H) + \sum_{s \in \mathcal{S} | w = 1 }r(s,a)b_H,
\end{equation}
where $r(s,a)$ is the original reward function for the POMDP. The discount factor $\gamma$ is the original discount factor of the POMDP. For the belief MDP $\mathcal{M}$, the expected value $V^{\pi}(q_0, b_{H,0})$ for policy $\pi$ starting from an initial state $(q_0, b_{H,0})$ is defined as:
\begin{equation}
	V^{\pi}= \sum_{t=0}^{\infty} \gamma^{t}r(q_t, b_{H,t}, a_t) =  \sum_{t=0}^{\infty} \gamma^{t} \mathbb{E}\left[r(s_t,a_t)| q_0, b_{H,0}, \pi\right],
\end{equation}
where the expectation is computed over $(q_t, b_{H,t}, w_t)$. The optimal fidelity selection policy maximizes the value in each belief state, i.e., $\pi^* =\arg\max_{\pi}V^{\pi}(q_0, b_{H,0}).$ We utilize the value iteration algorithm to solve the belief MDP and obtain the optimal fidelity selection policy.

\begin{remark}
For a similar setting, \cite{PG-VS:21m} theoretically established that the optimal policy under full state observability follows a threshold structure. This behavior appears in Fig.~\ref{ch_hil:fig:optimal_policy}, and their approach can be extended to formally characterize the optimal policy in the belief MDP.
\end{remark}
 
\section{Human Experiments}\label{ch_hil:sec:human_experiments}
In this section, we discuss the design of our human experiments conducted using Prolific (\url{www.prolific.com}).

We developed an underwater mine search experiment within the ROS framework using Gazebo models. We employed the ``UUV simulator"~\cite{7761080}, a comprehensive package that includes Gazebo plugins and ROS nodes for simulating unmanned underwater vehicles such as ROVs and AUVs.

In our experimental setup, underwater mines were placed randomly within a predefined area following a uniform distribution, and significant underwater vegetation was randomly introduced to further complicate mine detection.

\begin{figure}[ht]
 \centering
 \begin{subfigure}[b]{0.21\linewidth}
  \centering
\includegraphics[width=1\linewidth, height=1\linewidth, keepaspectratio]{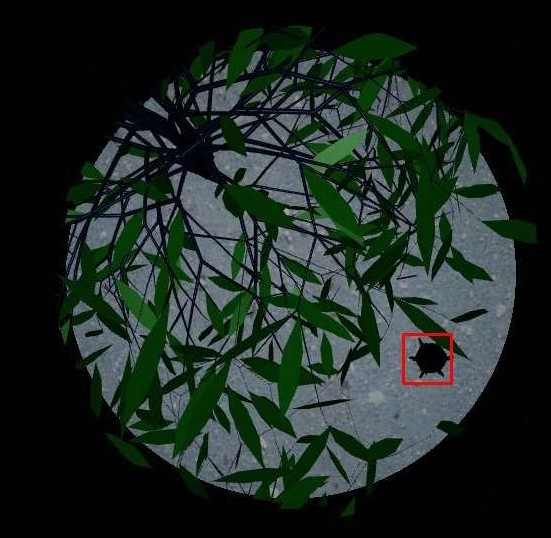}
\captionsetup{justification=centering} 
 \label{ch_hil:fig:mine1}
 \end{subfigure}
~
\begin{subfigure}[b]{0.21\linewidth}
  \centering
\includegraphics[width=1\linewidth, height=1\linewidth, keepaspectratio]{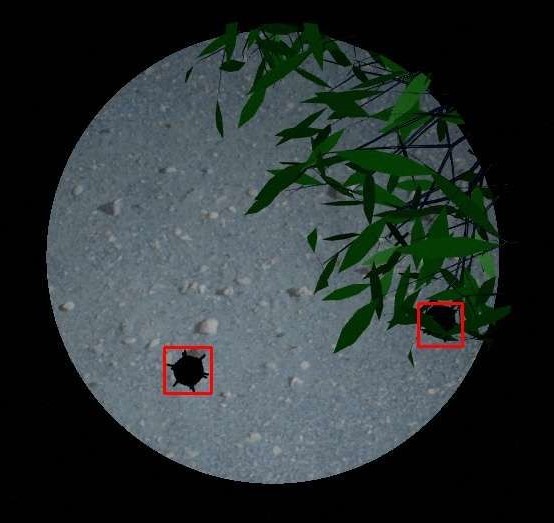}
\captionsetup{justification=centering} 
 \label{ch_hil:fig:mine2}
 \end{subfigure}
 ~
\begin{subfigure}[b]{0.21\linewidth}
  \centering
\includegraphics[width=1\linewidth, height=1\linewidth, keepaspectratio]{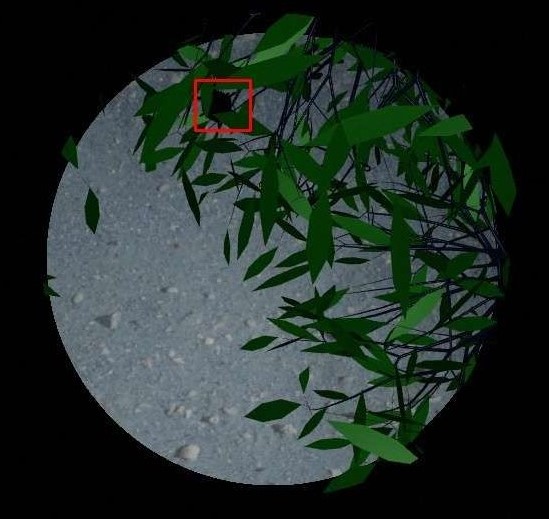}
\captionsetup{justification=centering} 
 \label{ch_hil:fig:mine3}
 \end{subfigure}
 ~
\begin{subfigure}[b]{0.2\linewidth}
  \centering
\includegraphics[width=1\linewidth, height=1\linewidth, keepaspectratio]{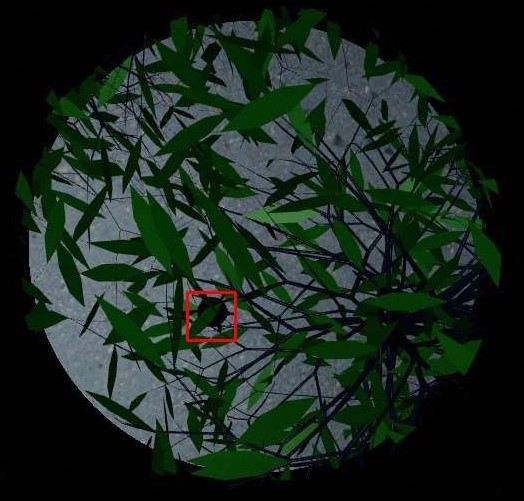}
\captionsetup{justification=centering} 
 \label{ch_hil:fig:mine4}
 \end{subfigure}

   \caption{\footnotesize Example image frames containing mines recorded from ROV. The mines' positions are highlighted within red bounding boxes in the frames.}
   \label{ch_hil:fig:mines}
\end{figure}

We captured underwater videos using an ROV that followed a predetermined circular trajectory through the environment. The ROV, equipped with a downward-facing camera, recorded images in complete darkness, illuminated solely by a single light directing its beam vertically downward. This setup ensured that only the area directly beneath the ROV was well illuminated. A total of $5600$ images were collected and organized into $56$ videos, each consisting of $100$ frames. Fig.~\ref{ch_hil:fig:mines} displays example frames where underwater mines are highlighted with red bounding boxes.

We conducted $5$ sets of experiments, with each group comprising $20$ participants. In each experiment, participants completed $8$ practice tasks followed by $48$ main tasks, each including both a primary and a secondary task. The following is a list of the experiments.

\begin{itemize}
    \item \textit{Experiment $1$:} This base experiment recorded data to train the IOHMM model and learn the optimal policy. The 48 videos were split evenly between normal and high fidelity, presented in blocks of four tasks. For half the participants, the sequence was arranged as $\{N, N, N, N, H, H, H, H, \ldots\}$, and for the other half, the order was reversed to mitigate ordering bias.    

    \item \textit{Experiment 2:} Participants chose the fidelity level for each task by pressing a key before task release. Task delegation was not permitted, with choices limited to normal or high fidelity.

    \item  \textit{Experiment 3:} A decision support system determined the fidelity level based on current operator performance and queue length. The system monitored performance, updated its belief using~\eqref{ch_hil:eq:belief_update}, and selected the optimal policy with two choices: normal and high fidelity. 

     \item \textit{Experiments 4 and 5:} Similar to Experiment $2$ and Experiment $3$, respectively, but participants were additionally provided with an option to delegate the task to the autonomous system.
\end{itemize}

\begin{figure}
\centering
 \includegraphics[width=0.8\linewidth, height=\linewidth, keepaspectratio]{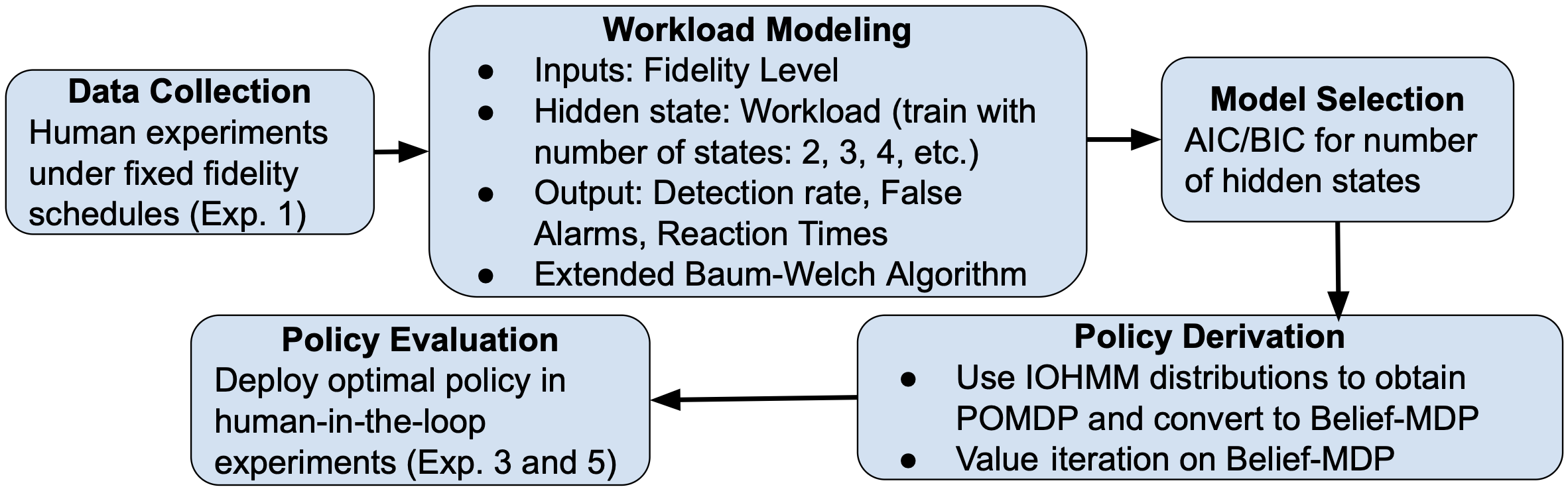}
 \caption{\footnotesize{Summary flowchart for the IOHMM training and policy derivation.}}
    \label{ch_hil:flowchart}
\end{figure}

Fig.~\ref{ch_hil:flowchart} illustrates the overall workflow, from human data collection through IOHMM training and policy synthesis.

Our experimental design is as follows:
\begin{itemize}
    \item \textbf{Independent variable}: Fidelity selection method (human-selected vs. optimal policy)
    \item \textbf{Dependent variable}: Cumulative reward (reflecting detection accuracy, false alarms, and queue penalties)
    \item \textbf{Control conditions}: Experiments 2 and 4 (manual fidelity selection)
    \item \textbf{Experimental conditions}: Experiments 3 and 5 (optimal policy selection)
    \item \textbf{Confounding control}: Participant screening \& task order counterbalancing (in Exp. 1) used to reduce bias
    \item All participants were recruited using identical criteria via Prolific and randomly assigned to experimental conditions to minimize systematic bias
    \item While covariates such as baseline reaction time or prior experience were not explicitly modeled, all participants received the same instructions, completed standardized practice trials, and had scores aggregated over 48 tasks to reduce intra-participant variability
\end{itemize}

\subsection{Methods and Participants' Instructions}\label{ch_hil:subsec:methods}
After receiving IRB consent (MSU IRB $\#9452$) from Michigan State University, 
we recruited $100$ participants via Prolific. Inclusion criteria required completion of at least $500$ prior studies and a $99\%$ approval rate. Participants received a base payment of $\$6$, with additional performance-based bonuses ranging from $\$0$ to $\$4$.

In the experiments, participants received detailed instructions for both the primary and secondary tasks to be performed simultaneously. 
They were instructed to maximize their scores based on their accuracy in identifying mines, with penalties for incorrect detections. Their bonus depended on their primary task performance and average secondary task response time, encouraging high scores on the primary task while maintaining fast secondary task responses.

Additionally, a dynamic queue displayed the number of videos to be processed, and the reward function was provided. Before the trials, each participant completed eight practice tasks. Depending on the experiment, either the fidelity level was selected by an optimal policy or chosen by the participants themselves.



\subsection{IOHMM Results}\label{subsec:IOHMM}

We use the data from Experiment $1$ to train IOHMM models with $2$, $3$, and $4$ hidden states. Table~\ref{ch_hil:table:aic_bic} shows the AIC and BIC values (normalized by the number of observation trajectories) for each model. Based on these criteria, we choose the IOHMM model with two hidden states, referred to as normal and high workload states.

\begin{table}[ht]
\centering
\begin{tabular}{|c|c|c|c|}
\hline
\textbf{Hidden States} & 2               & 3      & 4      \\ \hline
\textbf{AIC}           & \textbf{735.29} & 735.91 & 736.28 \\ \hline
\textbf{BIC}           & \textbf{736.39} & 737.85 & 739.26 \\ \hline
\end{tabular}
\caption{AIC and BIC values for model selection}
\label{ch_hil:table:aic_bic}
\end{table}

\begin{figure}[ht]
 \centering
 \begin{subfigure}[b]{0.4\linewidth}
  \centering
\includegraphics[width=1\linewidth, height=1\linewidth, keepaspectratio]{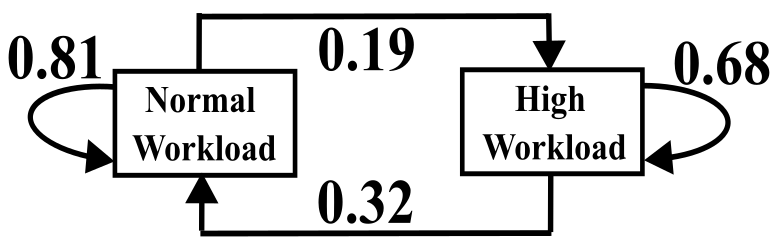}
\captionsetup{justification=centering} 
\caption{Normal fidelity}
 \label{ch_hil:fig:normal_fidelity_transition}
 \end{subfigure}
~
\begin{subfigure}[b]{0.4\linewidth}
  \centering
\includegraphics[width=1\linewidth, height=1\linewidth, keepaspectratio]{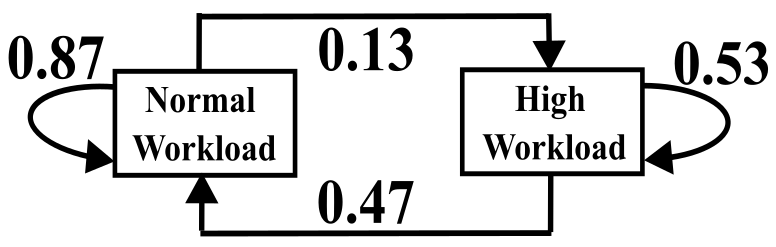}
\captionsetup{justification=centering} 
\caption{High fidelity}
 \label{ch_hil:fig:high_fidelity_transition}
 \end{subfigure}

   \caption{\footnotesize Workload transition distributions under (a) normal and (b) high fidelity.}
   \label{ch_hil:fig:state_transition}
\end{figure}
   
Fig.~\ref{ch_hil:fig:normal_fidelity_transition} and~\ref{ch_hil:fig:high_fidelity_transition} display the workload transition diagrams for the trained IOHMM model under normal and high fidelity, respectively. In high fidelity, where videos are slower, there is a higher probability of transitioning from a high to a lower workload state compared to normal fidelity. In contrast, under normal fidelity, the likelihood of transitioning from a normal to a high workload state is greater. For task delegation, the task is removed from the queue immediately, keeping the workload unchanged.

\begin{figure}[ht]
 \centering
 \begin{subfigure}[b]{0.47\linewidth}
  \centering
\includegraphics[width=1\linewidth, height=1\linewidth, keepaspectratio]{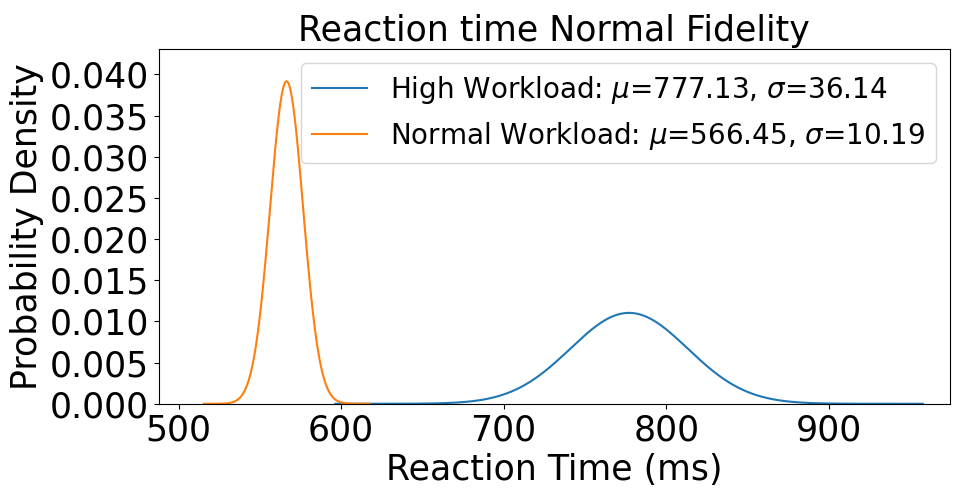}
\captionsetup{justification=centering} 
\caption{Normal fidelity}
 \label{ch_hil:fig:normal_fidelity_reaction}
 \end{subfigure}
~
\begin{subfigure}[b]{0.47\linewidth}
  \centering
\includegraphics[width=1\linewidth, height=1\linewidth, keepaspectratio]{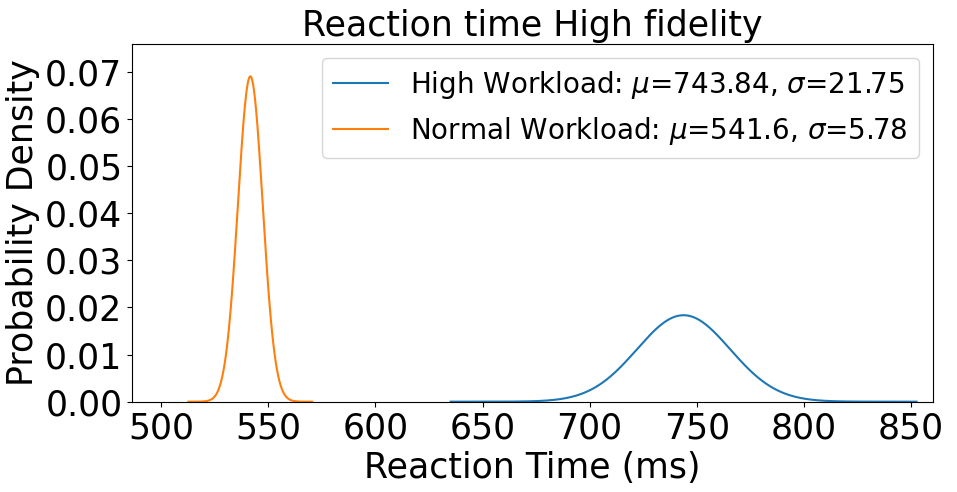}
\captionsetup{justification=centering} 
\caption{High fidelity}
 \label{ch_hil:fig:high_fidelity_reaction}
 \end{subfigure}

   \caption{\footnotesize Reaction time distributions under (a) normal and (b) high fidelity.}
   \label{ch_hil:fig:reaction}
\end{figure}
   
For observations $o^1$, $o^2$, and $o^3$, we learn a normal distribution with unknown mean and standard deviation for each state-action pair. Fig.~\ref{ch_hil:fig:normal_fidelity_reaction} and~\ref{ch_hil:fig:high_fidelity_reaction}  show the reaction time distributions for normal and high fidelity, respectively. State $0$ represents a normal workload state, while state $1$ represents a high workload state. The mean and variance of the secondary task reaction times are higher in the high workload state compared to the normal state.

\begin{figure}[ht]
 \centering
 \begin{subfigure}[b]{0.47\linewidth}
  \centering
\includegraphics[width=1\linewidth, height=1\linewidth, keepaspectratio]{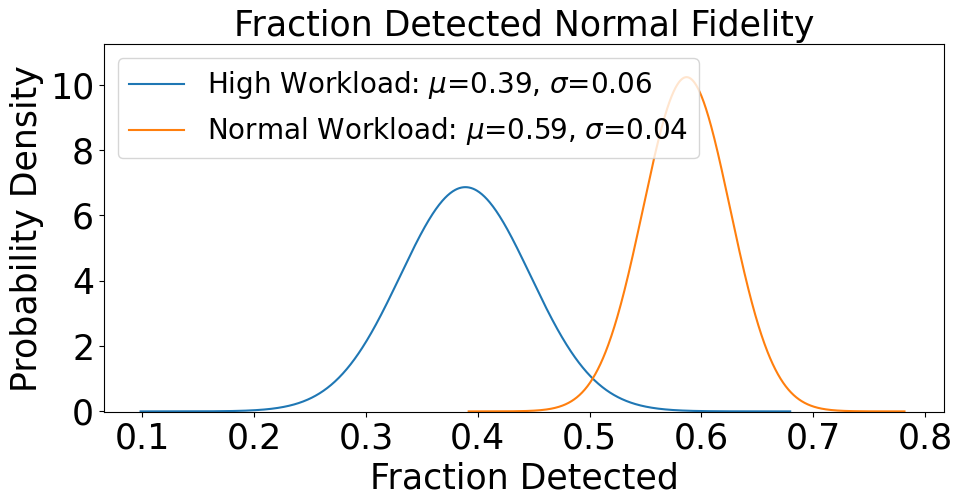}
\captionsetup{justification=centering} 
\caption{Normal fidelity}
 \label{ch_hil:fig:normal_fidelity_fraction}
 \end{subfigure}
~
\begin{subfigure}[b]{0.47\linewidth}
  \centering
\includegraphics[width=1\linewidth, height=1\linewidth, keepaspectratio]{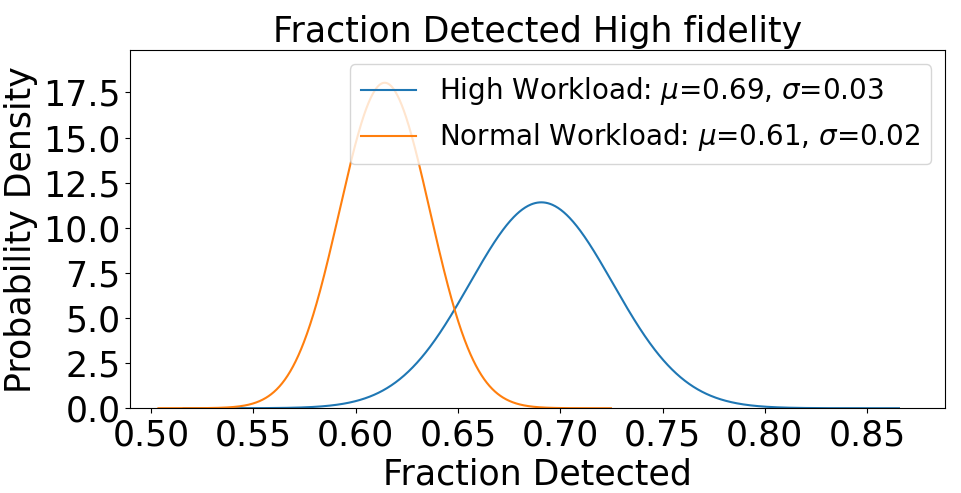}
\captionsetup{justification=centering} 
\caption{High fidelity}
 \label{ch_hil:fig:high_fidelity_fraction}
 \end{subfigure}

   \caption{\footnotesize Distributions for fraction of mines detected under (a) $N$  and (b) $H$.}
   \label{ch_hil:fig:fraction}
\end{figure}

Fig.~\ref{ch_hil:fig:normal_fidelity_fraction} and~\ref{ch_hil:fig:high_fidelity_fraction}  show the distributions for the fraction of detected mines in the primary task for normal and high fidelity, respectively. Under normal fidelity, the means of the distributions differ substantially between normal and high workload states, whereas the means are similar under high fidelity. This suggests that without considering queue penalties, using high fidelity (i.e., slower videos) could be advantageous during high workload conditions.


\begin{figure}[ht]
 \centering
 \begin{subfigure}[b]{0.47\linewidth}
  \centering
\includegraphics[width=1\linewidth, height=1\linewidth, keepaspectratio]{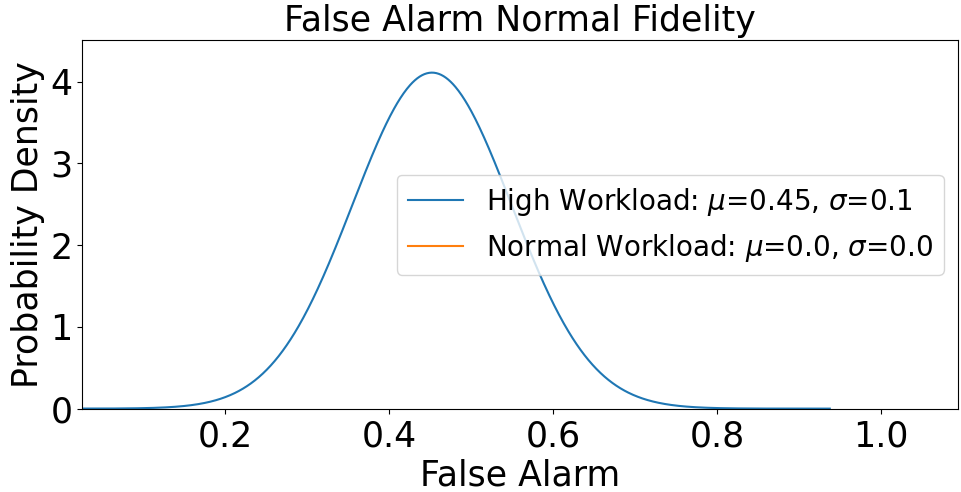}
\captionsetup{justification=centering} 
\caption{Normal fidelity}
 \label{ch_hil:fig:normal_fidelity_false}
 \end{subfigure}
~
\begin{subfigure}[b]{0.47\linewidth}
  \centering
\includegraphics[width=1\linewidth, height=1\linewidth, keepaspectratio]{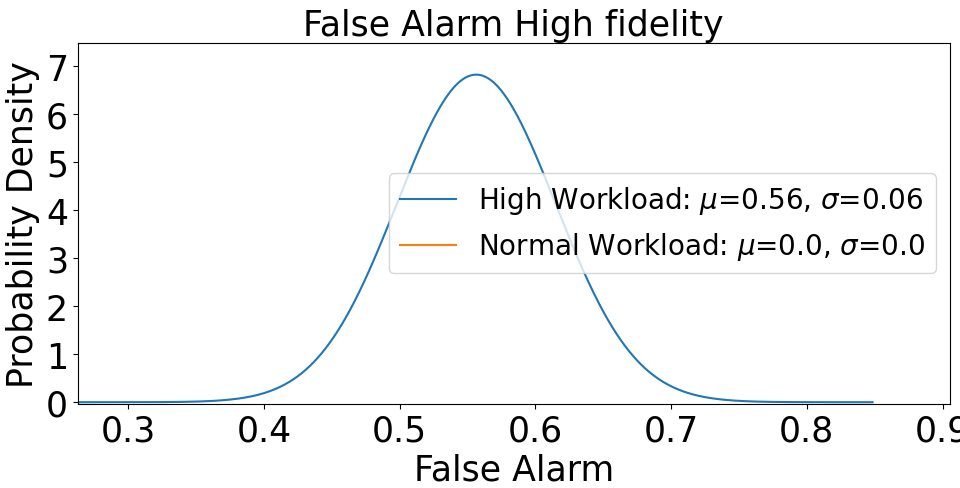}
\captionsetup{justification=centering} 
\caption{High fidelity}
 \label{ch_hil:fig:high_fidelity_false}
 \end{subfigure}
   \caption{\footnotesize Distributions for number of false alarms under (a) $N$ and (b) $H$.}
   \label{ch_hil:fig:false}
\end{figure}

Fig.~\ref{ch_hil:fig:normal_fidelity_false} and~\ref{ch_hil:fig:high_fidelity_false}  show the false alarm distributions for normal and high fidelity, respectively. In the normal workload state, zero false alarms were observed; therefore, the normal distribution is replaced by a Dirac delta function that outputs a probability of $1$ at $0$ false alarms and 0 elsewhere.

Finally, the initial distribution for the workload was determined as $[0.662,$ $\ 0.338$], where $0.662$ is the probability of starting in a normal workload state. To solve the POMDP, we discretize the distributions of the reaction time, fraction of mines detected, and the false alarms, with a step size of $25$ ms, $0.05$, and $0.5$, respectively.

\subsection{Optimal Policy}\label{ch_hil:optimal_policy}

Using the distributions from trained IOHMM, we convert the POMDP into a belief MDP as detailed in Sec.~\ref{ch_hil:subsec:mathematical_modeling}. We utilize the following reward function:
\begin{equation}\label{ch:hil:eq:reward}
    r(s,a) = \begin{cases}
        100o^1-30o^2-2q, & \text{for $a \in \{N, H\}$}, \\
        30 - 2(q-1), & \text{for $a=D$},
    \end{cases}
\end{equation}
where we assumed the accuracy of autonomous servicing (task delegation) to be just $30\%$. This choice of weights emphasizes mine detection accuracy while preventing excessive task delegation. We then used the value iteration algorithm to derive an optimal fidelity selection policy.

\begin{figure}[ht!]
 \centering
 \begin{subfigure}[b]{0.4\linewidth}
  \centering
\includegraphics[width=1\linewidth, height=1\linewidth, keepaspectratio]{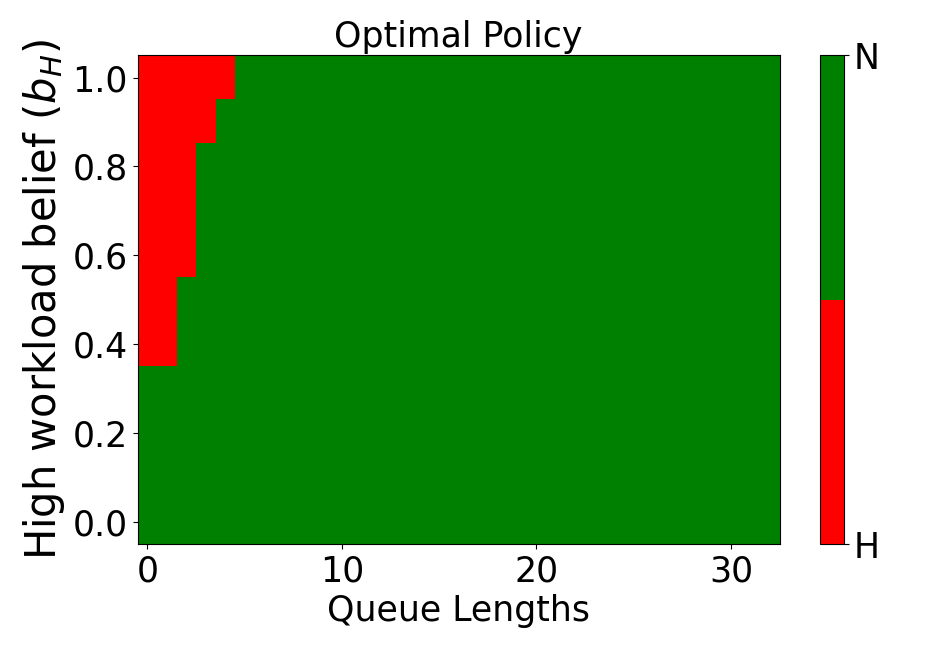}
\captionsetup{justification=centering} 
\caption{Task delegation unavailable}
 \label{ch_hil:fig:two_actions}
 \end{subfigure}
~
\begin{subfigure}[b]{0.4\linewidth}
  \centering
\includegraphics[width=1\linewidth, height=1\linewidth, keepaspectratio]{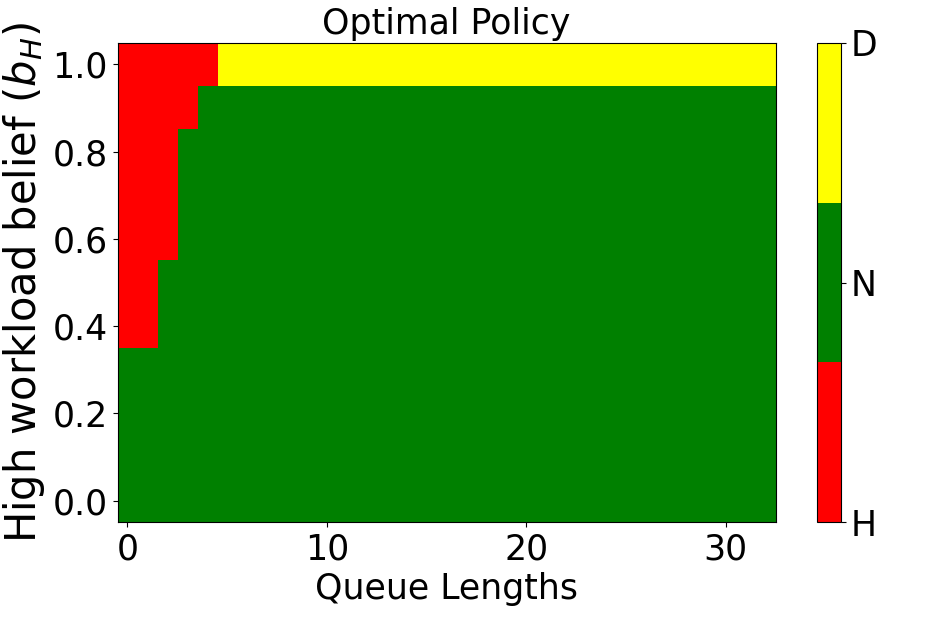}
\captionsetup{justification=centering} 
\caption{Task delegation available}
 \label{ch_hil:fig:three_actions}
 \end{subfigure}

   \caption{\footnotesize  Optimal fidelity selection policy where the action space (a) does not include task delegation  and (b) includes task delegation}
   \label{ch_hil:fig:optimal_policy}
\end{figure}

We derived two optimal policies: one with two actions and another with three actions that includes task delegation. Fig.~\ref{ch_hil:fig:two_actions} and~\ref{ch_hil:fig:three_actions} illustrate the optimal policy with two and three available actions, respectively. When the queue length is low and there is a high belief that workload is high, high-fidelity servicing is optimal; otherwise, normal fidelity is chosen. In the three-action policy, task delegation is selected only when the belief of being in a high workload state is almost certain (close to 1) and the queue length is large.

\section{Results and Discussion}\label{sec:resultsanddiscussion}
\begin{figure*}[ht]
 \centering
 \begin{subfigure}[b]{0.3\linewidth}
  \centering
\includegraphics[width=1\linewidth, height=1\linewidth, keepaspectratio]{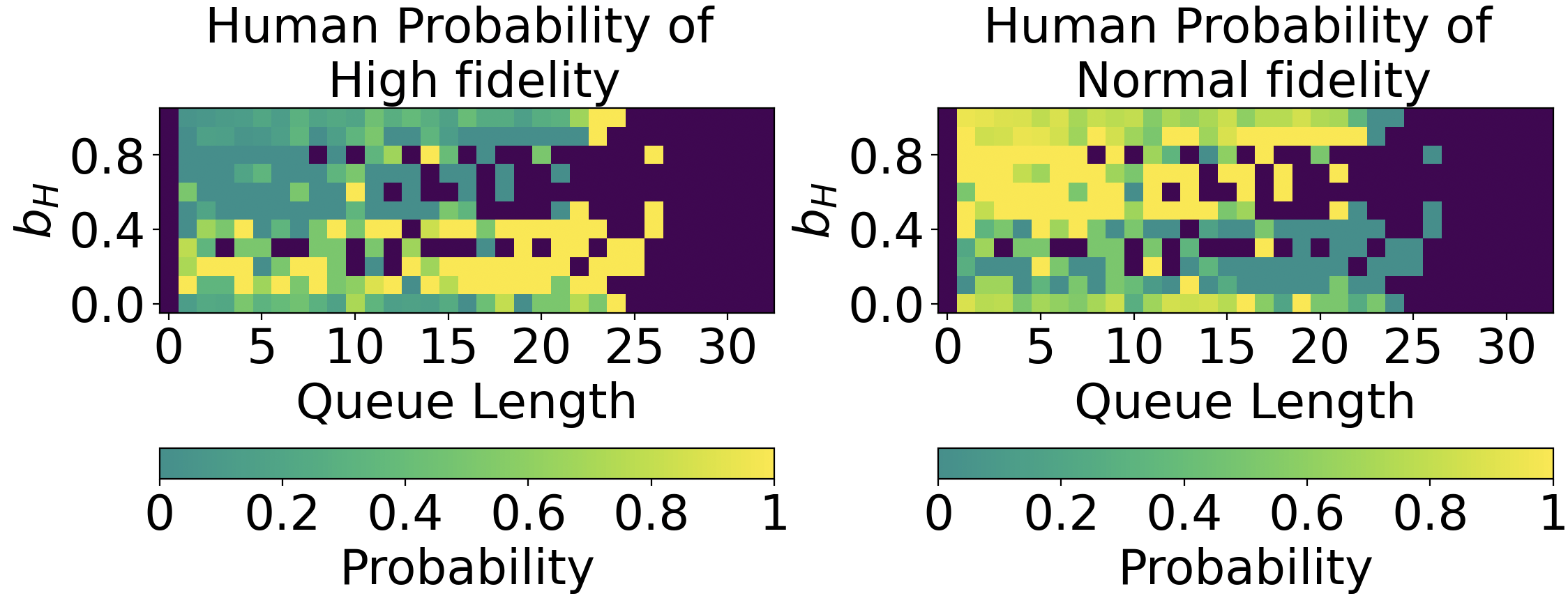}
\captionsetup{justification=centering} 
\caption{Experiment $2$ policy: Empirical human policy}
 \label{ch_hil:fig:human_without_skip}
 \end{subfigure}
~
\begin{subfigure}[b]{0.42\linewidth}
  \centering
\includegraphics[width=1\linewidth, height=1\linewidth, keepaspectratio]{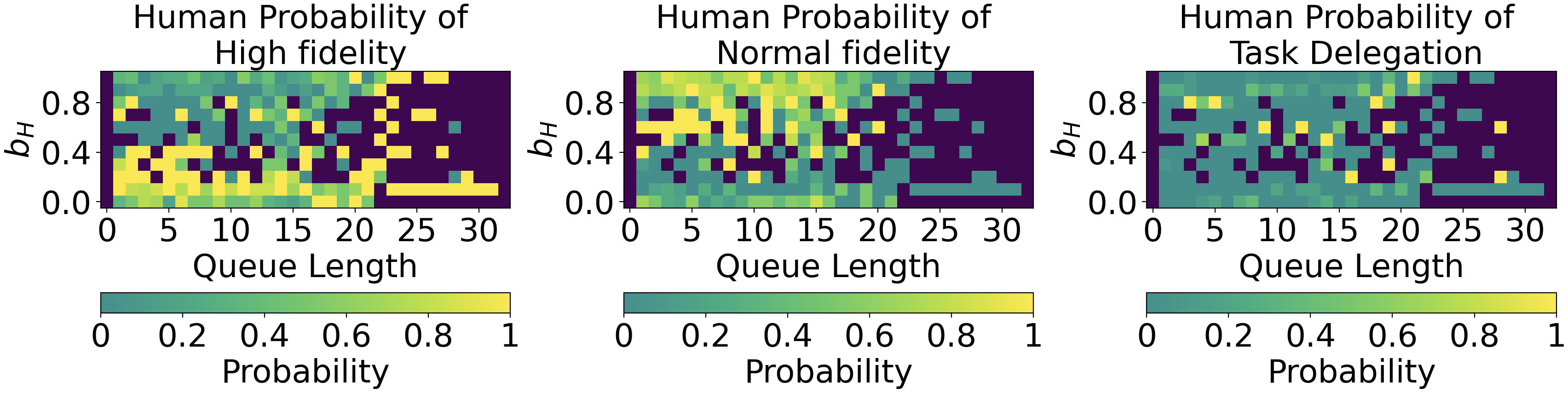}
\captionsetup{justification=centering} 
\caption{Experiment $4$ policy: Empirical human policy}
 \label{ch_hil:fig:human_with_skip}
 \end{subfigure}
~
\begin{subfigure}[b]{0.15\linewidth}
  \centering
 \includegraphics[width=1.05\linewidth, height=\linewidth, keepaspectratio]{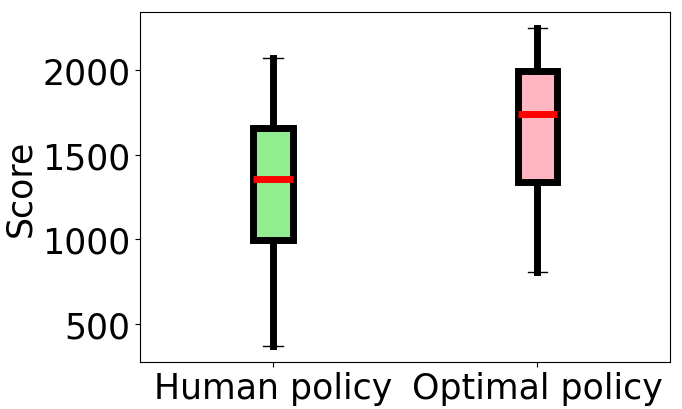}
 \caption{Exp. $2$ and $3$ box plot}
 \label{ch_hil:fig:boxplot_without_skip}
 \end{subfigure}

\begin{subfigure}[b]{0.3\linewidth}
  \centering
\includegraphics[width=1\linewidth, height=1\linewidth, keepaspectratio]{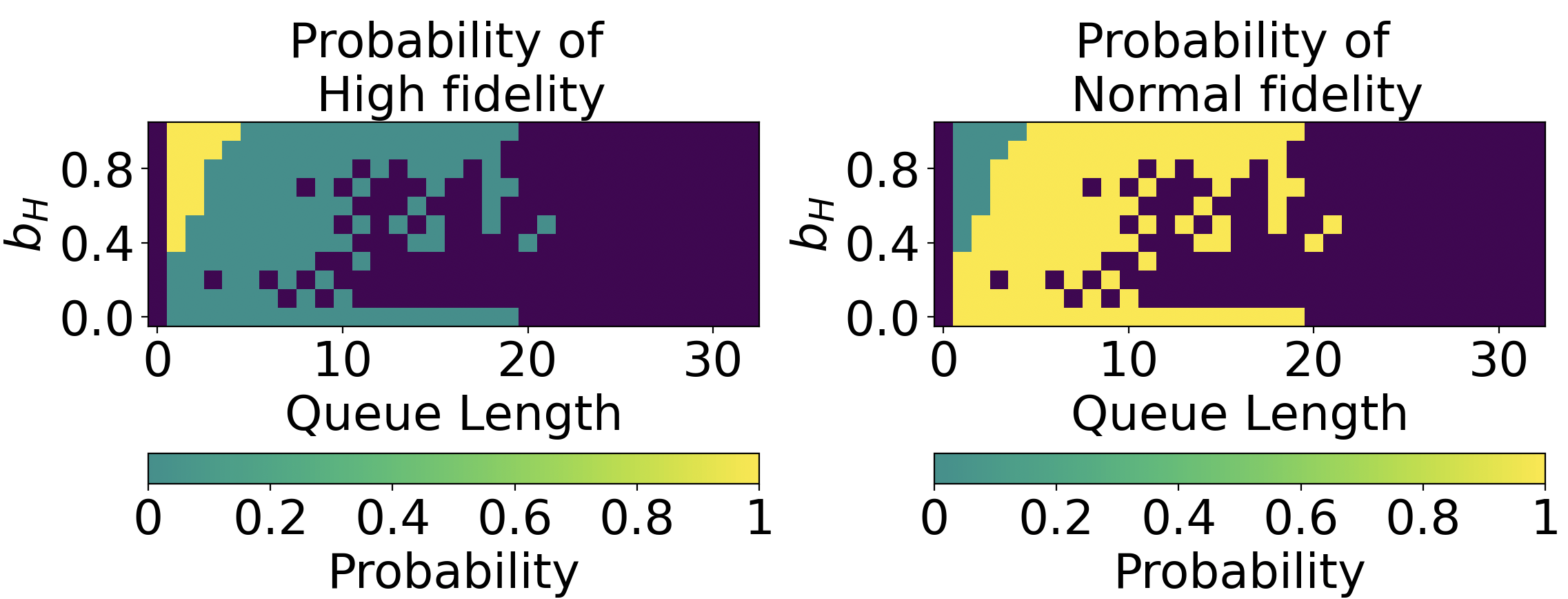}
\captionsetup{justification=centering} 
\caption{Experiment $3$ policy: Empirical optimal policy}
 \label{ch_hil:fig:optimal_without_skip}
 \end{subfigure}
 \begin{subfigure}[b]{0.42\linewidth}
  \centering
\includegraphics[width=1\linewidth, height=1\linewidth, keepaspectratio]{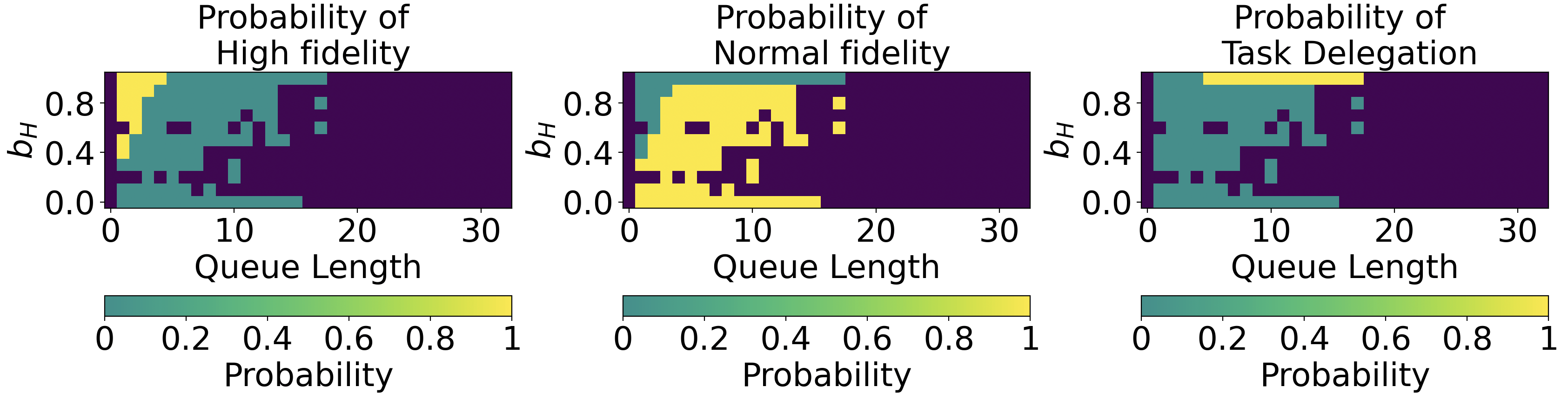}
\captionsetup{justification=centering} 
\caption{Experiment $5$ policy: Empirical optimal policy}
 \label{ch_hil:fig:optimal_with_skip}
 \end{subfigure}
 ~
\begin{subfigure}[b]{0.15\linewidth}
  \centering
 \includegraphics[width=1.05\linewidth, height=\linewidth, keepaspectratio]{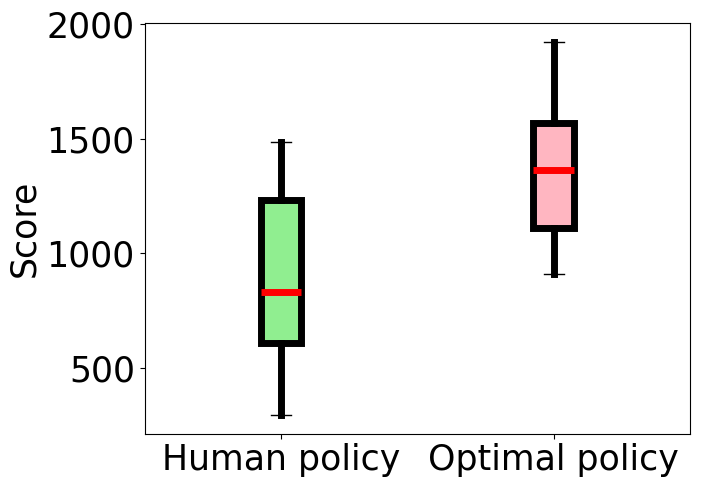}
 \caption{Exp. $4$ and $5$ box plot}
 \label{ch_hil:fig:boxplot_with_skip}
 \end{subfigure}
   \caption{\footnotesize  (a) and (b) shows empirical policies in Experiment $2$ and Experiment $4$, respectively. (c) and (d) shows the empirical policies from Experiment $3$ and Experiment $5$, respectively, that deploys the optimal policy. The plots (c) and (f) shows the box plots for the participants’ scores in Experiments $2$ and $3$, and Experiments $4$ and $5$, respectively. Within each box plot, the median is represented by the red horizontal line, while the lower and upper edges of the box signify the $25$th and $75$th percentiles, respectively. Whiskers extend to encompass the most extreme data points. }
   \label{ch_hil:fig:policy}
\end{figure*}

We now discuss the experimental results.
Figs.~\ref{ch_hil:fig:human_without_skip} and~\ref{ch_hil:fig:optimal_without_skip} illustrate the empirical policies from Experiment $2$ and Experiment $3$, respectively. In the plots, the two columns indicate the probabilities of choosing high fidelity (left) and normal fidelity (right). The dark purple regions represent unvisited portions of the state space during the experiment.

By comparing human policy with the optimal policy, we gain insights into human behavior. Fidelity shows minimal variation with queue length, suggesting that humans struggle to balance ongoing and pending tasks. They tend to favor high fidelity under low workload and normal fidelity under high workload, indicating a reluctance to switch actions. Participants who prioritize accuracy stick with high fidelity (low workload), while those focused on speed opt for normal fidelity (high workload). Overall, humans struggle to assess workload and performance, hindering effective action switching. In contrast, the optimal policy in Experiment $3$ effectively manages queue length, preventing the high-queue region from being explored.



Fig.~\ref{ch_hil:fig:boxplot_without_skip} presents the comparative box plots for the scores obtained in Experiments $2$ and $3$. Here, the score refers to the cumulative reward accrued by a participant over tasks, i.e.,  Score = $\sum_{t=1}^{48} r_t$, where $r_t$ denotes the reward obtained for servicing the task $t$ given by~\eqref{ch:hil:eq:reward}. Under the optimal policy, the performance improves with an increase of  $26.54\%$ in the average total score as compared to the human policy. 

Figs.~\ref{ch_hil:fig:human_with_skip} and~\ref{ch_hil:fig:optimal_with_skip} show the empirical policies from Experiments $4$ and $5$, respectively. The three columns of the plots show the probability of choosing high-fidelity, normal-fidelity, and task delegation, respectively. In addition to the conclusions drawn from Figs.~\ref{ch_hil:fig:human_without_skip} and~\ref{ch_hil:fig:optimal_without_skip}, 
it is noticeable that the human policy in Fig.~\ref{ch_hil:fig:human_with_skip} exhibits a somewhat random utilization of task delegation. This observation suggests that humans struggle with workload management and effective task delegation. Lastly, the deployed optimal policy in Fig.~\ref{ch_hil:fig:optimal_with_skip} keeps the queue length under check, and therefore, regions of higher queue lengths remain unvisited in the experiment.


Fig.~\ref{ch_hil:fig:boxplot_with_skip} shows the box plots for the participants' scores in Experiments $4$ and $5$. 
Under the optimal policy, we
observe an improved performance with an increase of $50.3\%$ in the average total score as compared to
the human policy. 

\begin{remark}
The average score in Experiment $5$ was marginally lower than in Experiment $3$. This likely reflects the relatively smaller reward for task delegation versus normal‐fidelity servicing. Although the optimal policy in Experiment $5$ may default to delegation under the presumption of diminished performance during high‐workload periods, some participants in Experiment $3$ demonstrate the ability to maintain high accuracy even under high workload.
\end{remark}

\begin{table}[ht]
\centering
\resizebox{\columnwidth}{!}{%
\begin{tabular}{|l|l|l|l|l|l|}
\hline
\textbf{\begin{tabular}[c]{@{}c@{}}Experiment \\ Comparison\end{tabular}} & 
\textbf{\begin{tabular}[c]{@{}c@{}}Group 1 \\ (Mean $\pm$ SD)\end{tabular}} & 
\textbf{\begin{tabular}[c]{@{}c@{}}Group 2 \\ (Mean $\pm$ SD)\end{tabular}} & 
\textbf{Cohen's $d$} & 
\textbf{\begin{tabular}[c]{@{}c@{}}Achieved \\ Power\end{tabular}} & 
\textbf{p-value} \\ \hline
Exp. 2 vs Exp. 3 & $1316.7 \pm 447$   & $1666.2 \pm 417.1$ & $0.81$ (large) & $0.704$  & $0.017$ \\ \hline
Exp. 4 vs Exp. 5 & $911.4 \pm 380.7$  & $1369.8 \pm 325.3$ & $1.29$ (very large) & $0.978$  & $0.001$ \\ \hline
\end{tabular}%
}
\caption{Post-hoc power analysis and statistical test results.}
\label{tab:power_analysis}
\end{table}

\textit{Statistical Analysis:} Table~\ref{tab:power_analysis} shows the results of the post-hoc power analysis and two-sample t-tests. These results indicate that the observed performance improvements under the optimal fidelity policies are statistically significant ($p < 0.05$), have large effect sizes (Cohen's $d$), and are supported by adequate to high power. While the power for Exp. 2 vs 3 ($0.704$) is slightly below the conventional $0.80$ threshold, the large effect size and statistically significant p-value ($p = 0.017$) suggest that the finding is robust and unlikely to be due to sampling variability.


\begin{table}[ht!]
\centering
\resizebox{0.9\columnwidth}{!}{%
\begin{tabular}{|cc|l|cc|}
\cline{1-2} \cline{4-5}
\multicolumn{2}{|c|}{\textbf{Transition Perturbation Sensitivity}} &  & \multicolumn{2}{c|}{\textbf{Reaction Time Noise Sensitivity}} \\ \cline{1-2} \cline{4-5} 
\multicolumn{1}{|c|}{\textbf{\begin{tabular}[c]{@{}c@{}}Transition \\ Perturbation (\%)\end{tabular}}} &
  \textbf{\begin{tabular}[c]{@{}c@{}}Reward Sensitivity\\ (abs. \% change)\end{tabular}} &
   &
  \multicolumn{1}{c|}{\textbf{\begin{tabular}[c]{@{}c@{}}Reaction Time Noise\\ $\sigma$ (ms)\end{tabular}}} &
  \textbf{\begin{tabular}[c]{@{}c@{}}Reward Sensitivity\\ (abs. \% change)\end{tabular}} \\ \cline{1-2} \cline{4-5} 
\multicolumn{1}{|c|}{0}                    & 0                     &  & \multicolumn{1}{c|}{0}                  & 0                   \\ \cline{1-2} \cline{4-5} 
\multicolumn{1}{|c|}{5}                    & 0.69                  &  & \multicolumn{1}{c|}{50}                 & 0.56                \\ \cline{1-2} \cline{4-5} 
\multicolumn{1}{|c|}{10}                   & 1.06                  &  & \multicolumn{1}{c|}{100}                & 1.49                \\ \cline{1-2} \cline{4-5} 
\multicolumn{1}{|c|}{20}                   & 3.76                  &  & \multicolumn{1}{c|}{150}                & 0.89                \\ \cline{1-2} \cline{4-5} 
\multicolumn{1}{|c|}{30}                   & 1.6                   &  & \multicolumn{1}{c|}{200}                & 1.22                \\ \cline{1-2} \cline{4-5} 
\multicolumn{1}{|c|}{40}                   & 4.83                  &  & \multicolumn{1}{c|}{250}                & 0.56                \\ \cline{1-2} \cline{4-5} 
\end{tabular}%
}
\caption{Sensitivity analysis under transition model perturbation and reaction time noise. Values indicate absolute percentage change in cumulative reward relative to the unperturbed baseline.}
\label{tab:sensitivity_analysis}
\end{table}

\textit{Sensitivity Analysis:} To evaluate the robustness of the proposed policy, we conduct a sensitivity analysis under two key sources of modeling uncertainty: (1) perturbations to the IOHMM workload transition probabilities and (2) additive Gaussian noise in reaction time, which directly impact workload estimation.

As shown in Table~\ref{tab:sensitivity_analysis}, the policy exhibits strong robustness across both conditions. Transition perturbations of up to $\pm40\%$ result in at most a 4.83\% absolute change in cumulative reward, while reaction time noise with standard deviations up to $250$ms leads to less than 1.5\% deviation. These findings demonstrate that the policy generalizes well and maintains stable performance despite significant model mismatch and observation noise.

\section{Conclusions and Limitations}\label{sec:conclusionsandlimitations}

We investigated optimal fidelity selection for human operators in a dual‐task visual search: participants simultaneously performed a primary mine‐search and a secondary visual task to estimate workload. We treated workload as a hidden state, modeled its dynamics via an IOHMM, 
and used it to solve a POMDP
for an optimal fidelity policy. Two experimental setups were tested: one permitting only normal or high fidelity, and another also allowing task delegation 
The optimal policy increased average scores by $26.54\%$ without delegation and $50.3\%$ with delegation, compared to a baseline where participants chose fidelity themselves.

This work has a few limitations. While reaction time is a widely used and standard measure of workload, it can also be influenced by other physiological factors. We acknowledge the absence of a subjective workload measure such as NASA-TLX in our study. Including such surveys could strengthen validation and is a direction for future work.
Additionally, POMDPs require extensive training data, which is why we limit our model to a small, discrete state space and action space. Consequently, scalability becomes a challenge when applying this approach to larger state and action spaces.
To address this, future work could leverage scalable approaches such as belief compression (e.g., point-based value iteration, low-rank approximations)~\cite{pineau2006anytime, dave2024approximate}, parametric policy learning, or neural approximators trained via deep reinforcement learning~\cite{kimura1997reinforcement}.

 \bibliographystyle{ieeetr} 
 \bibliography{references}

\end{document}